\documentclass{iaus}
\usepackage{graphicx}
\usepackage{subfigure}

\title[Reverberation Mapping at MDMO] 
{Reverberation Mapping Results from MDM Observatory}

\author[Kelly~D.~Denney et al.]  
{Kelly~D.~Denney$^1$, B.~M.~Peterson$^1$, R.~W.~Pogge$^1$,
M.~C.~Bentz$^2$, C.~M.~Gaskell$^3$, T.~Minezaki$^4$,
C.~A.~Onken$^5$, S.~G.~Sergeev$^{6,7}$, M.~Vestergaard$^{8,9}$}

\affiliation{$^1$The Ohio State University, 140 West 18th Avenue,
Columbus, OH 43210, USA; denney@astronomy.ohio-state.edu $^2$University
of California at Irvine $^3$University of Texas at Austin $^4$University
of Tokyo $^5$Mount Stromlo Observatory, Australia $^6$Crimean
Astrophysical Observatory $^{7}$Isaak Newton Institute of Chile
$^{8}$Steward Observatory $^{9}$University of Copenhagen}

\pubyear{2009}
\volume{267}  
\jname{Co-Evolution of Central Black Holes and Galaxies}
\editors{B.M.\ Peterson, R.S.\ Somerville, \& T.\ Storchi-Bergmann, eds.}
\begin{document}

\maketitle

\begin{abstract}
We present results from a multi-month reverberation mapping campaign
undertaken primarily at MDM Observatory with supporting observations
from around the world. We measure broad line region (BLR) radii and
black hole masses for six objects.  A velocity-resolved analysis of the
H$\beta$ response shows the presence of diverse kinematic signatures in
the BLR.

\end{abstract}



Reverberation mapping takes advantage of the presence of a time delay or
lag, $\tau$, between continuum and emission line flux variations
observed through spectroscopic monitoring campaigns to infer the radius
of the broad line region (BLR) and, subsequently, the central black hole
mass in type 1 AGNs.  The primary goal of this campaign was to obtain
either new or improved H$\beta$\ reverberation lag measurements for
several relatively low luminosity AGNs.  Using cross correlation
techniques to measure the time delay between the mean optical continuum
flux density around 5100\AA\ and the integrated H$\beta$\ flux, we
determine the H$\beta$\ lags and black hole mass measurements listed in
Columns 2 and 3 of Table 1, respectively.  Column 4 tells if this
measurement is new, an improvement meant to replace a previous, less
reliable measurement, or simply an additional measurement not used to
replace a previous value.  The complete results from this study are
currently being prepared for publication (Denney et al., in
preparation).  A subsequent velocity-resolved analysis of the H$\beta$\
response shows that three of the six primary targets demonstrate
kinematic signatures (Column 5) of infall, outflow, and non-radial
virialized motions (see \cite[Denney et al. 2009]{Denney09}).

%
%
\begin{table}[h]
\centering
\caption{Mean H$\beta$ Lags and Black Hole Masses}
\begin{tabular}{lcccc}

Object & $\tau_{\rm cent}$(days) & M$_{\rm BH}(\times 10^6 {\rm M}_\odot)$ & Data Use & Kinematic Signature\\
\hline
NGC\,3227 & $3.75^{+0.76}_{-0.82}$ & $7.63^{+1.62}_{-1.72}$ & improvement & infall\\ 
NGC\,3516 & $11.68^{+1.02}_{-1.53}$ & $31.7^{+2.8}_{-4.2}$ & improvement & outflow \\ 
NGC\,5548 & $12.40^{+2.74}_{-3.85}$ & $44.2^{+9.9}_{-13.8}$ & add'l measurement  & virial\\ 
Mrk\,290 & $8.72^{+1.21}_{-1.02}$ & $24.3^{+3.7}_{-3.7}$ & new & --- \\
Mrk\,817 & $14.04^{+3.41}_{-3.47}$ & $61.9^{+15.0}_{-15.3}$ & add'l measurement & --- \\
NGC\,4051 & $1.87^{+0.54}_{-0.50}$ & $1.73^{+0.55}_{-0.52}$ & improvement & --- \\
\hline
\end{tabular}

\end{table}

\end{document}